# *Defects in ultrathin copper nanowires*

Jeong Won Kang*, Jae Jeong Seo, Ki Ryang Byun, and Ho Jung Hwang

Semiconductor Process and Device Laboratory, Department of Electronic Engineering, Chung-Ang University, 221 HukSuk-Dong, DongJak-Ku, Seoul 156-756, Korea

We have performed atomistic simulations for cylindrical multi-shell (CMS)-type Cu nanowires containing defects. Our investigation has revealed some physical properties that have not been detected in previous studies that have considered defect-free nanowires. Since the vacancy formation energy is lowest in the core of a CMS-type nanowire, a vacancy formed in the outer shell of a CMS-type nanowire naturally migrates toward the core. The maximum of the formation energy of an adhered atom on the surface of a CMS-type nanowire was modeled using a 16-11-6-1 nanowire. The formation energy of an adhered atom decreased when the diameter of the CMS-type nanowire was either above or below the diameter of the peak energy maximum. This investigation found three recombination mechanisms for the vacancy–adhered atom pairs: (i) by direct recombination, (ii) by a kick-in recombination, and (iii) by a ring recombination. Vacancy formation energy calculations show that an onion-like cluster with a hollow was formed, and molecular dynamics simulations for various CMS-type nanowires found that vacancies migrated towards the core. From these, we obtained basic information on the formation of hollow CMS-type metal nanowires (metal nanotubes) [Y. Oshima, *et al*., Phys. Rev. B 65, 121401 (2002)].

## 1. Introduction

Recently, ultra-thin metal nanowires have aroused growing interest in condensed matter physics; for example, Takayanagi's group has fabricated ultra-thin gold [1-3] and platinum [4] nanowires. Many theoretical studies on ultra-thin nanowires have been done using atomistic simulations for several metals, and these have simulated straight-line uniform ultra-thin nanowires containing helical multi-shell structures, such as Ag [5], Al [6], Au [7-10], Ti [11], Cu [12], and Zr [13]. In addition to studies on novel helical structures of ultra-thin nanowires, the melting behavior of ultra-thin nanowires has been investigated for Pb [14], Au [15], Cu [16], and Ti [17]. The compression of cylindrical multi-shell (CMS)-type Au nanowires [18] and the tensile testing of CMS-type Cu nanowires [19] have also been performed. The resonance of ultra-thin Cu nanobridges was investigated using a classical molecular dynamics (MD) simulation [20]. However, further investigations, in areas such as non-linear ultra-thin nanowires, funnel-shaped nanowires, and defects in nanowires need to be made in order to understand the physical properties of nanowires, and for the successful application of nanowires to nanoscale devices. Defects in nanowires have been found in previous work using MD simulations [12,21,22]. Figure 5(b) in Ref. [21] shows vacancy–adhered atom pairs on the surface of a decagonal nanowire, Fig. 3 in Ref. [12] shows a vacancy on the surface of a CMS-type nanowire, and Fig. 4(c) in Ref. [22] also shows a vacancy–adhered atom pair on the surface of a CMS-type nanopillar. However, a study on defects in CMS-type nanowires has not been made until now.

Defects in carbon nanotubes (CNTs) have been investigated [23-25], and these investigations were valuable for the understanding of the physical properties of CNTs. Although these previous studies have disclosed novel structural information and melting behaviors of ultra-thin metal nanowires, our knowledge of ultra-thin metal nanowires is still quite limited. Since ultra-thin nanowires have CMS-type structures, this investigation focuses on defects, such as vacancies, adhered atom, and vacancy-adhered atom pairs, in CMS-type Cu nanowires. This study on defects in CMS-type nanowires provides further physical information on the structural properties of CMS-type nanowires.

## 2. Methods

For the Cu-Cu interactions, we have used a many-body potential function of the second-moment approximation of the tight-binding scheme [26] that has already been tested in nanoclusters [27-30] and nanowires [31,32], among others. The potential reproduces many basic properties of crystalline and non-crystalline bulk phases and surfaces [33], and provides a good insight into the structures and thermodynamics of metal clusters [34,35].

Each shell of the CMS-type nanowires is composed of a circular folding {111} sheet, and the core is the atomic strand [3,8,12]. Kondo and Takayanagi [3] introduced the notation $n$ - $n'$- $n''$- $n'''$, to describe a nanowire consisting of coaxial tubes with $n$, $n'$, $n''$, $n'''$ helical atom rows (where $n > n' > n'' > n'''$). In this paper, we use the notation by Kondo and Takayanagi to describe a CMS-type nanowire. In previous papers [11-13], CMS-type nanowires have had a relationship in which the number of atomic rows between shells differs by five (*i.e.*, n - n´ = 5).

In this work, we used the MD and the steepest descent (SD) methods. The MD simulations used the same MD methods as used in our previous works [21,36]. The MD code used the velocity Verlet algorithm, a Gunsteren-Berendsen thermostat to keep constant temperature, a periodic boundary condition along the wire axis, and neighbor lists to improve computing performance [37].

## 3. Results and Discussion

We investigated the defects in five nanowires shown in the Tables. The supercells of the CMS-type nanowires were composed of thirty layers along the wire axis, and Table 1 shows the number of atoms of the supercells. Figure 1 shows the spreading sheet of a shell that includes a vacancy (V) site and an adhered atom (A) site. While the positions of vacancies are clearly understood, the adhered atom positions need further explanation. The adhered atom site on the {111} sheet is at the center of mass of a triangle. However, the adhered atom sites on the shell made by a circular folding of the {111} sheet are always at the center of mass of a lozenge. This is because of the curvature of the shell in both the MD and the SD simulations. When an atom is at the center of mass of a triangle, the atoms along the arc of the shell, which are labeled 1 and 2 in Fig. 1, become more distant, and the adhered atom moves to the center of mass of the lozenge, as indicated by the arrow in Fig. 1. For considering a vacancy in CMS-type nanowires, one atom is omitted in the core or shells of the nanowires, and then the structures of the nanowire are fully relaxed using the SD method. For an adhered atom, one atom is added to the surface of the nanowire, and then the structures of the nanowire are fully relaxed using the SD method. For the vacancy–adhered atom pair, one adhered atom is added to the surface of the structure obtained from the above vacancy calculation, and then the relaxed structure is found using the SD method. After full relaxation, the formation energy of defects ($E_f$) can be defined as

$$E_f = E_{Dt} - \frac{N_{Dt}}{N_N} E_N, \qquad (3\text{-}1)$$

where $E_{Dt}$ is the total cohesive energy of the nanowire including a defect, $E_N$ is the total cohesive energy of a perfect nanowire, $N_{Dt}$ is the number of atoms composing the nanowire including the defect, and $N_N$ is the number of atoms composing the perfect nanowire.

### 3.1 Vacancy formation energy

Table 2 shows the mono-vacancy formation energy of each shell of a CMS-type nanowire. The vacancy formation energies of the outer shells of CMS-type nanowires are slightly less than those on the {111} surface, and are similar to each other regardless of the diameter of the CMS-type nanowire. For the inner shells of the CMS-type nanowires, as the diameter of the inner shell increases, the vacancy formation energy increases. In all cases, except for the 11-6-1 nanowire, the vacancy formation energies of the inner shells are always higher than those of the outer shell. Figure 2 shows graphs for values in Table 2 in order to explain the trend in vacancy formation energies. Figures 2(a) and 2(b) shows the vacancy formation energies of shells as a function of a diameter for three CMS-type nanowires and the vacancy formation energies of cores and outer shells versus the reciprocal of the diameter of a CMS-type nanowire for a vacancy, respectively. As the diameter of the CMS-type nanowire increases, the vacancy formation energy of its core decreases rapidly toward zero. Therefore, these results imply that a vacancy is most frequently found in the core of a CMS-type nanowire, and that a vacancy is also formed more easily in the outer shell than in the inner shell of a CMS-type nanowire having more than three shells.

### 3.2 Formation energy of the adhered atom

We also calculated the formation energy of the adhered atom in a CMS-type nanowire, and the results are shown in Table 3. The smaller the diameter of a CMS-type nanowire, the lower the formation energy of the adhered atom. This means that as the diameter of the CMS-type nanowire decreases, the surface of the CMS-type nanowire more easily adsorbs an atom. Figure 3 shows the formation energy *versus* the reciprocal of the diameter of a CMS-type nanowire for an adhered atom. The maximum formation energy of an adhered atom was observed for the 16-11-6-1 nanowire. As the diameter of the CMS-type nanowire increased above the formation energy maximum, the formation energy of the adhered atom decreased and approached that of an adhered atom on the {111} surface. As the diameter of the CMS-type nanowire decreased below that of the formation energy maximum, the formation energy of the adhered atom decreased rapidly.

### 3.3 Formation energy of the vacancy-adhered atom pair

In previous studies, MD simulations have sometimes shown the adhered atom on the surface [21,22]. Since these MD simulations were performed assuming perfect CMS-type nanowires, an adhered atom on the surface implies that a vacancy is included in the nanowire. Therefore, we performed calculations on a simple vacancy-adhered atom (V-A) pair assuming a nearest neighbor distance between the vacancy and the adhered atom. The V-A pair can be of two types, as shown in Fig. 4. One is a V-A pair along the wire axis denoted by Type I in Fig. 4, and the other is a V-A pair denoted by Type II in Fig. 4, where the adhered atom is at the adhered site discussed in Fig. 1. Table 4 shows the formation energy of the V-A pairs. The formation energy of Type I pairs is higher than that of Type II pairs, because the adhered site of a Type II pair is more stable than that of the adhered site of a Type I pair, as discussed in Fig. 1. In both cases, as the diameter of the wire increases, the formation energy increases correspondingly.

### 3.4 Recombination of the vacancy–adhered atom pair

Furthermore, one can think that the recombination mechanism of Type I pairs is different from that of Type II. Figures 5 and 6 show snapshots of the recombination mechanism for a 5-1 nanowire with Type I V-A pairs, and a 6-1 nanowire with Type II V-A pairs, respectively. The adhered atom in Figs. 5 and 6 is indicated by the symbol *I*. Parts *(a)* to *(d)* in Figs. 5 and 6 correspond to steps *(a)* to *(d)* shown in Fig. 7, which outlines the energy diagram of V-A pair. In the case of a Type I pair, the adhered atom directly recombines with the vacancy. Owing to the presence of the adhered atom, the distance between Atoms 1 and 2 in Fig. 5 becomes longer, with the distance between Atoms 3 and 4 also becoming slightly longer, and so the adhered atom moves into the vacancy over the recombination energy barrier. This occurs in the case of a direct recombination of the V-A pair. In the case of a Type II pair, atoms (Atom 1 and Atom 2 in Fig. 6), which are on both sides of a vacancy site along the arc of a CMS-type nanowire, move closer together, and then the adhered atom leaves the adhered site for recombination to take place. Finally, the adhered atom forces Atom 2 into the vacancy site. This is the case of a kick-in recombination of the V-A pair. While the adhered atom of a Type I pair plays a trigger role in the direct recombination of the V-A pair, the vacancy of the Type II pair takes the lead in the kick-in recombination of the V-A pair.

The formation energy barrier of a V-A pair in Figure 7 is the sum of the formation energy and the recombination energy barrier of the V-A pair. Using the recombination mechanisms of the V-A pair shown in Figs. 5 and 6, we calculated the recombination energy barrier of a V-A pair. In the case of direct recombination of the V-A pair, Atom *I* was displaced toward the vacancy, and then the structure was relaxed using the SD method, as long as the lattice position of *I* was fixed. In the case of a kick-in recombination of a V-A pair, Atom 2 of Fig. 6 was displaced toward the vacancy along the arc of the CMS-type nanowire, and then the structure was relaxed using the SD method, as long as the lattice position of Atom 2 was fixed. The calculated values were very low, as shown in Table 5. In the case of the Type I V-A pair, as the diameter of the CMS-type nanowire increased, the distance between Atoms 3 and 4 in Fig. 5(a) became shorter. Therefore, the direct recombination energy barrier of a V-A pair increased. In the case of a Type II V-A pair, as the diameter of the CMS-type nanowire increased, the distance between Atoms 1 and 2 in Fig. 6 became shorter. Therefore, the kick-in recombination energy barrier of the V-A pair decreased.

**3.5 MD simulations**

The results obtained from the SD simulations were not sufficient to provide an understanding of the properties of defects. Therefore, we performed MD simulations for the V-A pairs. The results obtained from the MD simulations showed the same results as those obtained from the SD simulations, but also they showed results for various cases that were not obtained from the SD simulations, which are discussed below. Figure 8 shows the results of an MD simulation at 70 K for a 11-6-1 nanowire with a V-A pair. In this case, the recombination of the V-A pair did not occur, but the vacancy was exchanged with an atom in the inner shell, *R*, in Figs. 8(a) and 8(b), and an adhered atom, *I*, in Fig. 8(a) diffused on the surface. As discussed in Table 1, since the vacancy formation energy of the inner shell is lower than that of the outer shell, the vacancy migrated from the outer shell to the inner shell. Therefore, we can see that the vacancy migrates towards the lower energy state. Since the formation energy of a vacancy is lowest at the core, the vacancy in Fig. 8(a) migrated to the core, as shown in Fig. 8(b). The ring of the inner shell wrapping the vacancy transformed from a six-atom ring into a five-atom ring, and one interstitial site was generated, as shown on the left side in Fig. 8(b). After the atom, *R,* of the inner shell moved to the vacancy of the outer shell, the five-atom ring was formed at that point in the inner shell. The core atom surrounded by the five-atom ring bore the stress, and then a chain-like movement of the core atoms towards the left side was achieved. Therefore, the nanowire included one vacancy and one interstitial site. The nanowire became bent, centering on the position of the vacancy. Oshima *et al*. [4] showed a single-shell Pt nanotube without an atomic strand in the core. Therefore, we can infer by analogy that when CMS-type nanowires have serial vacancies in the core, they can then be considered as CMS-type nanotubes. Some MD simulations have also shown that vacancies in the inner shell or the outer shell migrate towards the core. These are similar to the hollow spherical clusters optimized by the SD method after MD simulation. We have performed MD simulations to try to obtain ball-like metal structures, such as seen in $C_{60}$. However, we did not obtain any ball-like structures, but we did obtain a $Cu_{80}$ cluster that was composed of two shells consisting of pentagons and hexagons, as shown in Figs. 9(a) and 9(b). This is similar to the multi-layered fullerene cages, which are also called "bucky onions" [38-42]. The outer-ring and the inner-ring consist of 12 and 6 atoms, respectively. Figures 9(a) and 9(b) show the outer shell composed of 60 atoms and the inner shell composed of 20 atoms, respectively. The inside of Fig. 9(b) is vacant and can contain just two atoms. If this structure were extended along the z-axis (the axis of atomic strand in core), then a caged Cu nanotube would result. Although our results are limited by the simulation conditions, methods, and interatomic potentials, this result shows an aspect, the absence of a central atomic strand, that is related to the results of Oshima *et al*. [4].

Figure 10 shows the different result of MD simulations at 300 K. In this case, an atom from the inner shell, *R*, moves to the vacancy site. However, the vacancy recombines with an atom from the outer shell, *S*. The adhered atom, *I*, forces atom *S* into recombination with the vacancy. This mechanism is called a ring recombination mechanism and is different from the kick-in mechanism. Figure 11 also shows a different type of kick-in mechanism. In the case of Fig. 6, the adhered atom, *I*, ejects atom *S*, which is contained in the same outer-ring that includes the vacancy, into the vacancy. This is called a kick-in mechanism induced by a vacancy. However, in the case of Fig. 11, the adhered atom, *I*, ejects atom *S*, which that is in the same outer-ring as the adhered atom, into the vacancy. Therefore, this is called a kick-in mechanism led by an adhered atom.

Since the vacancy formation energy at the core of CMS-type nanowires is the lowest, a vacancy in the inner or outer shell migrates to the core. Therefore, during the growth or formation of CMS-type nanowires, as many vacancies in the core region may be generated, these structures may be found as CMS-type nanotubes, as described in Reference [4]. When a V-A pair is formed on the surface of a CMS-type nanowire, if the recombination of the V-A pair does not occur rapidly, then the vacancy will migrate towards the core, and the adhered atom will migrate towards another adhered site on the surface. The V-A pair recombination processes can be classified by three mechanisms: (1) direct recombination, (2) a kick-in recombination, and (3) a ring recombination. The kick-in mechanism is of two types: those induced by a vacancy, and those led by the adhered atom.

## 4. Conclusion and Future Work

Previous theoretical papers [5-19] have focused on structural predictions and the thermal behavior of ultra-thin metal nanowires. This study on CMS-type nanowires containing defects has revealed some physical properties that were not detected in previous works that considered defect-free-nanowires. The vacancy formation energy was found to be the lowest in the core of CMS-type nanowires. The maximum formation energy of an adhered atom exists in a 16-11-6-1 nanowire. When the diameter of the CMS-type nanowire is above or below the diameter of the 16-11-6-1 nanowire with the formation energy peak maximum, then the formation energy of an adhered atom decreases. This investigation showed two types of V-A pairs and three V-A pair recombination mechanisms. From the vacancy formation energy, an onion-like cluster with a hollow is formed, and MD simulations for various cases show that the vacancies migrate towards the core. From our studies, we obtained basic information on the formation of CMS-type metal nanotubes.

However, we could not provide the reason that the vacancy formation energy of the CMS-type nanowire is lowest in the core and has a maximum near the surface. Therefore, our future works will include this topic and a study on the properties of optimal metal nanotubes

**Tables**

Table 1. Numbers of atoms composing perfect CMS-type nanowires.

| CMS | 5-1 | 6-1 | 11-6-1 | 16-11-6-1 | 21-16-11-6-1 |
|---|---|---|---|---|---|
| Number of atoms | 180 | 210 | 540 | 1020 | 1650 |

Table 2. Vacancy formation energy of CMS-type nanowires.

| CMS | Vacancy formation energy (eV) | | | | |
|---|---|---|---|---|---|
| | Shell including vacancy | | | | |
| | Core | 1st | 2nd | 3rd | 4th |
| 5-1 | 0.313286 | 0.968486 | | | |
| 6-1 | 0.510764 | 0.821064 | | | |
| 11-6-1 | 0.169626 | 0.620626 | 0.807626 | | |
| 16-11-6-1 | 0.086741 | 1.042741 | 1.320741 | 0.775741 | |
| 21-16-11-6-1 | 0.054385 | 0.929385 | 1.219385 | 1.266385 | 0.801385 |
| Vacancy formation energy on {111} surface | | | | | 0.941788 |

Table 3. Formation energy of an adhered atom on the surface of CMS-type nanowires.

| CMS | Formation energy of an adhered atom on the surface (eV) |
|---|---|
| 5-1 | 0.190414 |
| 6-1 | 0.334136 |
| 11-6-1 | 0.557374 |
| 16-11-6-1 | 0.825259 |
| 21-16-11-6-1 | 0.791615 |
| {111} surface | 0.582211 |

Table 4. Formation energy of the vacancy – adhered atom (V-A) pair on the surface of CMS-type nanowires.

| CMS | Formation energy of a V-A pair on the surface (eV) | |
| --- | --- | --- |
| | Type I | Type II |
| 5-1 | 1.2246 | 1.1549 |
| 6-1 | 1.4054 | 1.2105 |
| 11-6-1 | 1.5540 | 1.3940 |
| 16-11-6-1 | 1.8080 | 1.6040 |
| 21-16-11-6-1 | 2.0520 | 1.8600 |

Table 5. Recombination energy barrier of V-A pairs on the surface of CMS-type nanowires.

| CMS | Recombination energy barrier (eV) | |
| --- | --- | --- |
| | Type I | Type II |
| 5-1 | 0.0021 | 0.0495 |
| 6-1 | 0.0582 | 0.0454 |
| 11-6-1 | 0.0698 | 0.0431 |
| 16-11-6-1 | 0.0809 | 0.0411 |
| 21-16-11-6-1 | 0.0820 | 0.0405 |

**Figures**

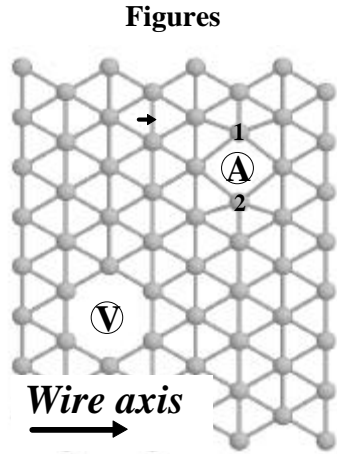

Figure 1. Spreading sheet of a shell including a vacancy (V) and an adhered atom (A) sites.

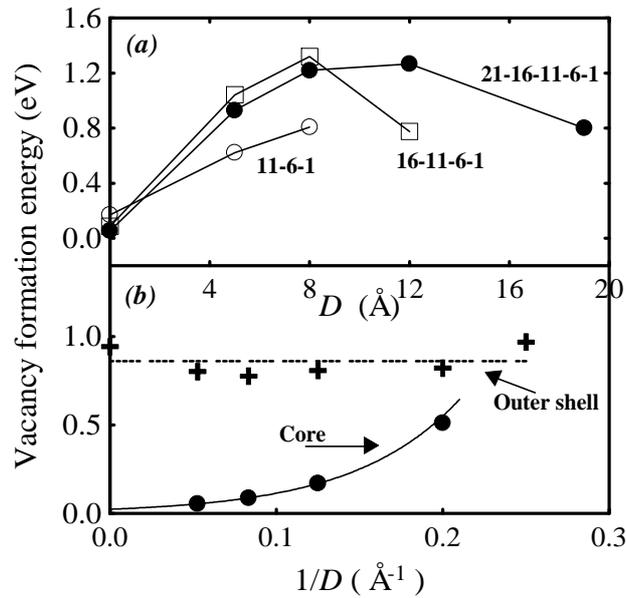

Figure 2. (a) The vacancy formation energies of shells as a function of a diameter for three CMS-type nanowires and (b) the vacancy formation energies of cores and outer shells versus the reciprocal of the diameter of a CMS-type nanowire for a vacancy.

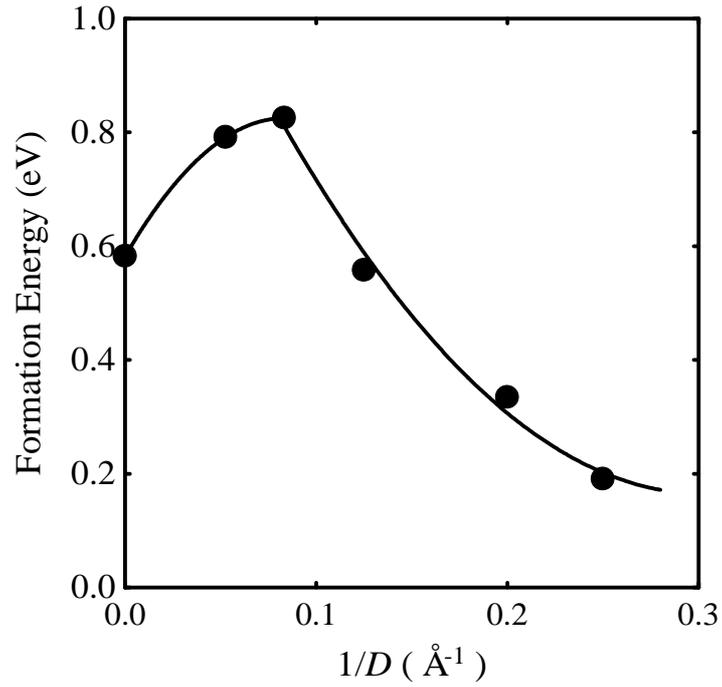

Figure 3. Formation energy *versus* the reciprocal of the diameter of a CMS-type nanowire for an adhered atom.

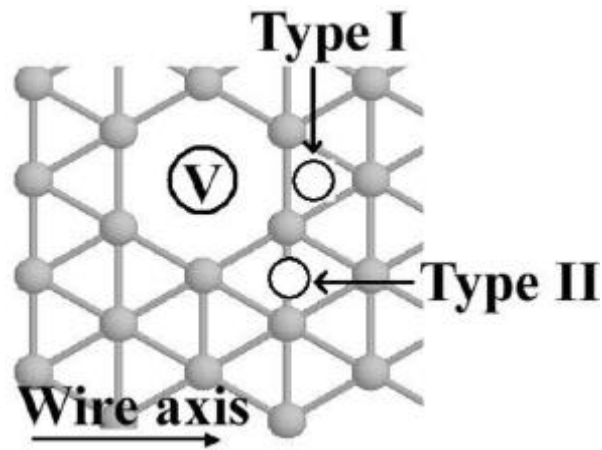

Figure 4. Two types of the vacancy – adhered atom (V-A) pairs, where V is the vacancy, and the circle indicates the adhered atom sites.

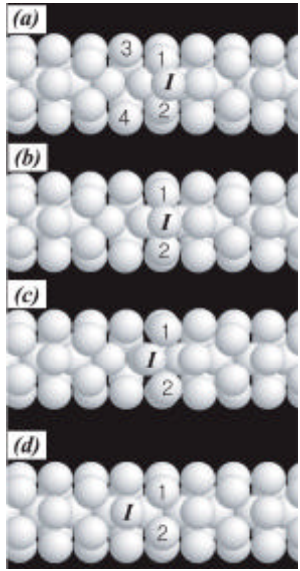

Figure 5. Snapshots of the direct recombination of a V-A pair (Type I) in a 5-1 nanowire using the SD method. The label *I* denotes the adhered atom.

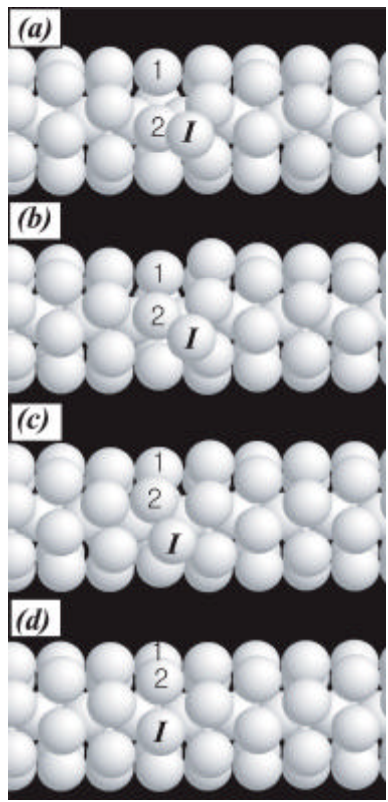

Figure 6. Snapshots of the kick-in recombination of a V-A pair (Type I) in a 6-1 nanowire using the SD method. The label *I* denotes the adhered atom.

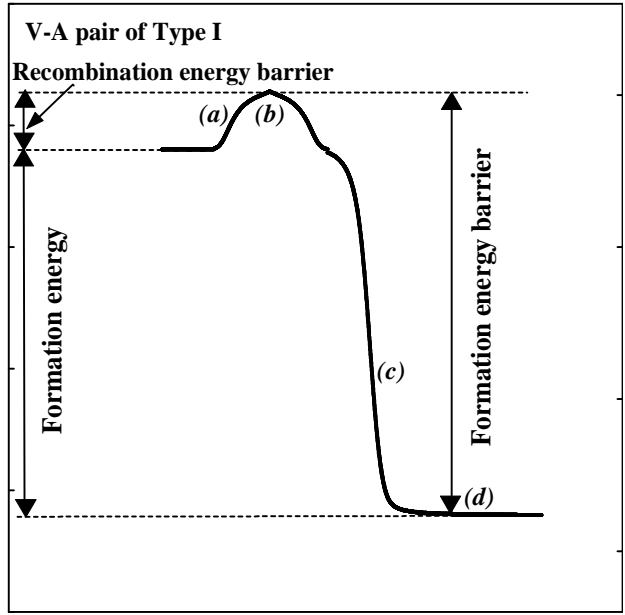

Figure 7. Energy diagram related to a V-A pair (Type I).

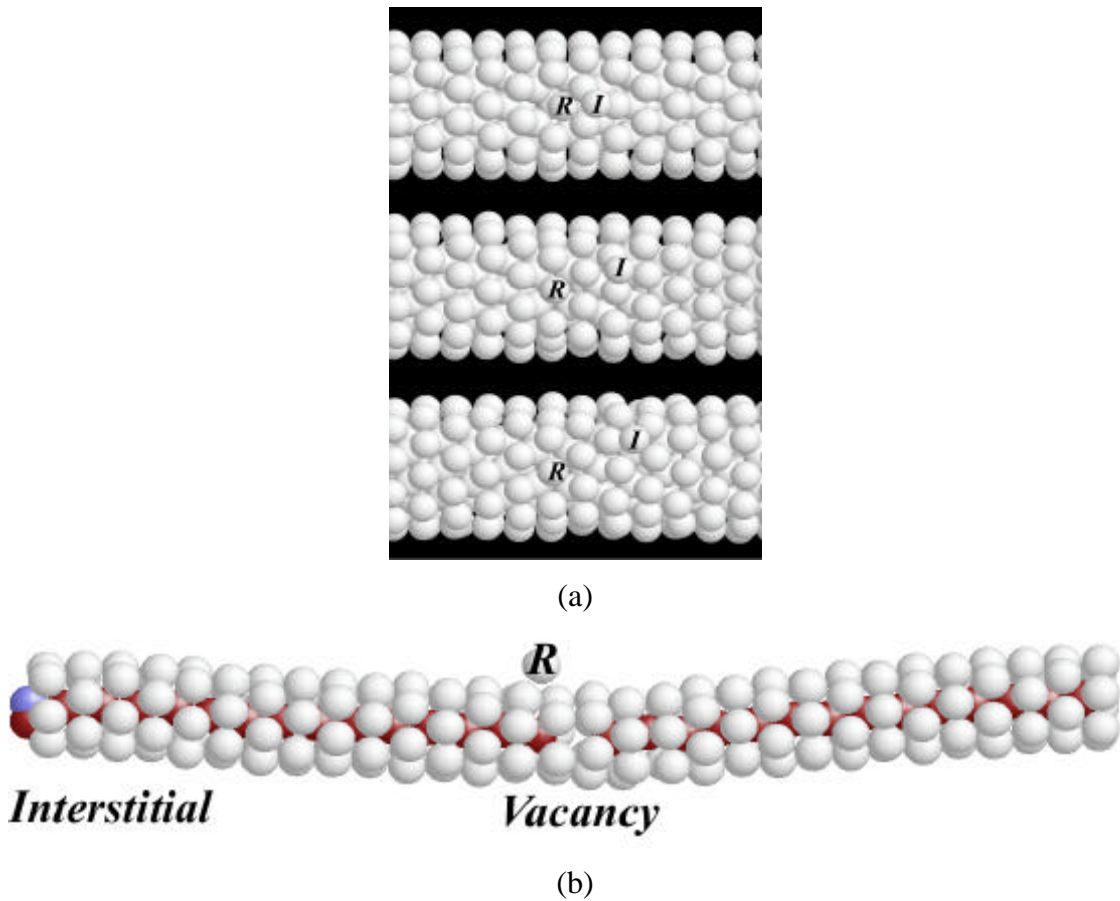

(a)

(b)

Figure 8. Result of the MD simulations at 70 K for a 11-6-1 nanowire with a V-A pair (Type I). (a) Snapshots of the vacancy and the adhered atom migrations, where *I* denotes the adhered atom and *R* denotes the atom moved from the inner shell to the vacancy of the outer shell. (b) Structure of the inner shell and the core including a vacancy and an interstitial site.

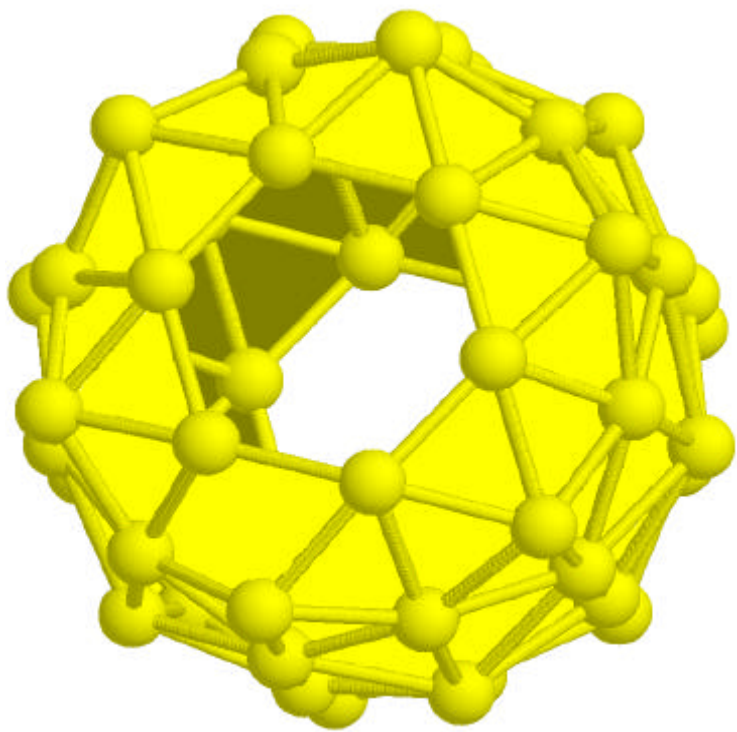

(a)

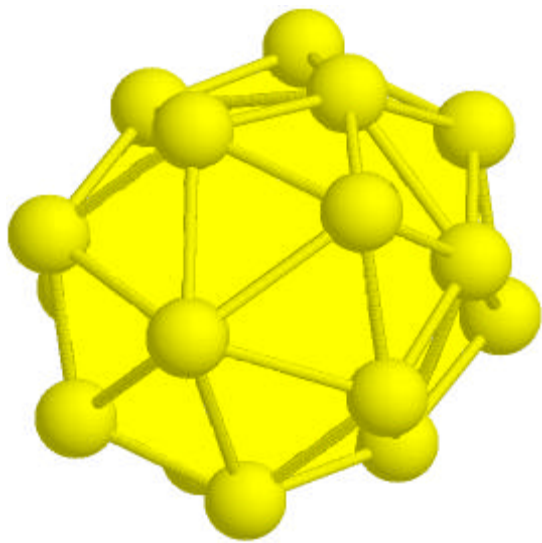

(b)

Figure 9. Structures of the shells of an onion-like $Cu_{80}$ cluster. (a) The outer shell, and (b) the inner shell with a hollow that possesses two atoms.

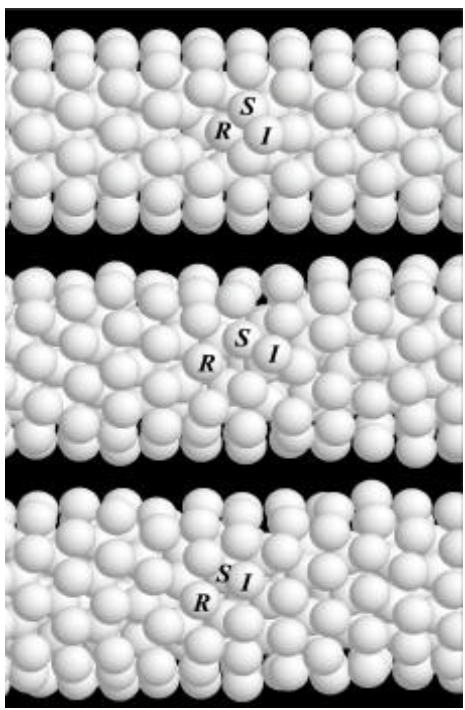

Figure 10. Snapshots of the MD simulation at 300 K for a 11-6-1 nanowire with a V-A pair (Type I). The labels *I*, *R*, and *S* denote the adhered atom, the atom moved from the inner shell to the vacancy of the outer shell, and the atom of the outer shell recombined with the vacancy in the inner shell, respectively.

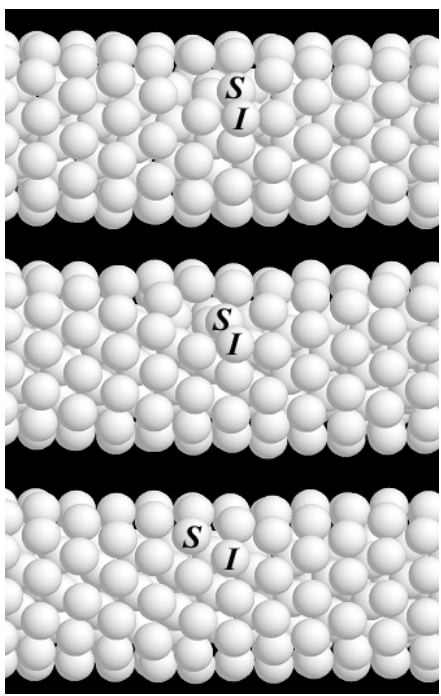

Figure 11. Snapshots of the MD simulation at 70 K for a 11-6-1 nanowire with a V-A pair (Type II). The labels *I* and *S* denote the adhered atom and the atom of the outer shell recombined with the vacancy in the outer shell, respectively.